\documentstyle[aps,twocolumn]{revtex}
\input epsf

\begin{document}


\draft

\title{Angular momentum at the black hole threshold}

\author{Carsten Gundlach}
\address{Max-Planck-Institut f\"ur Gravitationsphysik
(Albert-Einstein-Institut), Schlaatzweg 1, 14473 Potsdam, Germany}

\date{26 Nov, revised 21 Dec 1997}

\maketitle


\begin{abstract}

Near the black hole threshold in phase space, the black hole mass as a
function of the initial data shows the ``critical scaling''
$M\simeq C(p-p_*)^\gamma$, where $p$ labels a family of initial data,
$p_*$ is the value of $p$ at the threshold, and the critical exponent
$\gamma$ is universal for a given matter model. The black hole charge
$Q$ obeys a similar law. To complete the picture, we include angular
momentum as a perturbation. For the black hole angular momentum $\vec
L$ we find the oscillating behavior $\vec L \simeq {\rm Re}
\left[(\vec A + i \vec B) (p-p_*)^{\mu+i\omega}\right]$.  The
assumptions of the calculation hold for $p=\rho/3$ perfect fluid
matter, and we calculate $\mu\simeq
0.799$ and $\omega
\simeq 0.231$.

\end{abstract}



\paragraph{Introduction}

There has been much interest recently in critical phenomena at the
black hole threshold. Take a smooth one-parameter family of smooth,
asymptotically flat initial data for general relativity (with matter)
that crosses the black hole threshold, and fine-tune the parameter, so
as to get close to the black hole threshold. For many families, one
can make arbitrarily small black holes this way. As discovered by
Choptuik
\cite{Choptuik}, the black hole mass near the threshold then scales as
$M\simeq C (p-p_*)^\gamma$, where $p$ is the parameter of the initial
data, $p_*$ is its value at the black hole threshold, and $\gamma$,
the ``critical exponent'' is a universal constant common to all
one-parameter families, while $C$ depends on the family.

In a nutshell, critical phenomena arise because general relativity,
for a given matter model and symmetry, admits a ``critical
solution''. By this we understand a strong-field solution that is an
attractor of codimension one sitting on the black hole threshold, and
whose basin of attraction is the black hole threshold. Perturbatively
this means that the critical solution has precisely one unstable mode,
and that it decays into a black hole, or into flat space, depending on
the sign of that mode. Critical solutions found so far are either
static or periodic in time (type I), or continuously or discretely
self-similar (type II). Both can arise in the same matter model
\cite{ChoptuikChmajBizon,BradyChambersGoncalves}. Here we limit
attention to type II, which gives rise to the power-law scaling of the
mass. For a review of critical phenomena, see
\cite{Gundlach_critreview}. 

If a small amount of electric charge is present in the initial data,
the black hole charge $Q$ obeys a similar power law as the mass $M$,
with a new critical exponent $\delta$
\cite{GundlachMartin,HodPiran_charge}. Black holes are completely
characterized by their mass $M$, electric charge $Q$ and angular
momentum $\vec L$. To complete the picture, here we consider initial
data with small angular momentum. Our calculation assumes the
existence, for a given matter model, of a spherically symmetric
self-similar solution with exactly one unstable perturbation mode,
including nonspherical perturbations.

Recently, we have shown for the first time that a spherically
symmetric critical solution, namely that for perfect fluid matter with
equation of state $p=\rho/3$, has no unstable {\it non}spherical
perturbation modes \cite{Gundlach_codim}. The present abstract
calculation therefore applies to this particular matter model, and
gives to a quantitative prediction for it.

\paragraph{Notation}

The calculation we give here is hardly more than an application of
dimensional analysis.  Therefore we set up a compact notation that
hides all technical details specific to general relativity, in order
to bring out the essential scaling argument. Let $g_{\mu\nu}$ be the
spacetime metric in the coordinates $(\tau,x,\theta,\varphi)$, with
$\theta$ and $\varphi$ the usual Euler angles, and let surfaces of
constant $\tau$ be spacelike.  If the conformal metric $\tilde
g_{\mu\nu} \equiv e^{2\tau} g_{\mu\nu}$ is independent of $\tau$, then
the spacetime is homothetic, or continuously self-similar (CSS). In
this case, $\tau$ plays the double role of being the time coordinate
and of being the negative logarithm of spacetime scale, while $x$ is
dimensionless. A simple example of such a coordinate system is
$\tau=-\ln(-t)$ and $x=r/(-t)$, where $t$ and $r$ are a standard
choice of time and radial coordinate.

Let $Z(\tau,x,\theta,\varphi)$ stand for $\tilde g_{\mu\nu}$ plus a
suitable set of conformally rescaled matter variables. (For perfect
fluid matter, for example, these are $\tilde
\rho = e^{-2\tau} \rho$ for the density, and $\tilde u_\mu = e^{\tau}
u_\mu$ for the four-velocity.)  $Z(\tau,x,\theta,\varphi)$ then
symbolizes a solution of the field equations, while
$Z_0(x,\theta,\varphi)$ symbolizes Cauchy data. (To make this notion
precise, we have to transform from $\tilde g_{\mu\nu}$ to suitably
rescaled Cauchy data $\tilde h_{ij}$ and $\tilde K_{ij}$.) Finally,
let $Z(\tau,x,\theta,\varphi)=Z_*(x)$ stand for the critical solution,
which is both spherically symmetric and CSS. (We overlook a trivial
$\theta$ dependence, namely the factor $\sin^2\theta$ in
$g_{\varphi\varphi}$, for the sake of notation.)

For a solution that is a general perturbation of a self-similar,
spherically symmetric background solution $Z_*(x)$, we can make the ansatz
\begin{eqnarray}
\label{ansatz}
\nonumber
Z(\tau,x,\theta,\varphi) \simeq && Z_*(x) + \sum_{l=0}^\infty
\sum_{m=-l}^l \sum_{n=0}^\infty 
\\ 
&& {\rm Re}\left[ C_{lmn}(p) \,
e^{\lambda_{ln} \tau} \, f_{lmn}(x,\theta\varphi)\right].
\end{eqnarray}
Because the background is spherically symmetric, we can decompose all
linear perturbations into spherical harmonics labeled by $l$ and
$m$. Because it is independent of $\tau$, the perturbation equations
of motion separate, and the perturbations can be assumed to have a
simple exponential dependence on $\tau$, where the separation
constants $\lambda_{lmn}$ and the mode functions $f_{lmn}$ are
complex.  We assume here that the perturbation spectrum is countable
and discrete, and label it by $n$ for each value of $l$ and
$m$. Because $Z$ is real, the $\lambda_{lmn}$ and $f_{lmn}$ must be
real or form complex conjugate pairs. Because of spherical symmetry,
the $\lambda_{lmn}$ are in fact the same for all $m$, and are
therefore labelled $\lambda_{ln}$. The $C_{lmn}$, also complex, are
the free parameters of the generic perturbation. Here we want to
consider one-parameter families of initial data labelled by a
parameter $p$. The dependence of the solutions on $p$ is then a
dependence of the free constants $C_{lmn}$ on $p$.  The $f_{lmn}$
correspond to rescaled metric and matter perturbations in a suitable
gauge, such as Regge-Wheeler gauge, or in a gauge-invariant
framework. We have addressed the technical details of gauge-fixing
elsewhere
\cite{Gundlach_codim}. 

\paragraph{Review of mass scaling}

We now consider the time evolution of initial data that start close to
the spherically symmetric self-similar critical solution, so that the
expansion (\ref{ansatz}) applies initially. The critical solution
contracts to a curvature singularity at $\tau=\infty$. As the
singularity is approached from the past, with $\tau$ increasing, a
perturbation grows (decays) if the real part of its $\lambda$ is
positive (negative). By definition, a critical solution has precisely
one unstable mode. Being unique, this unstable mode must be
spherically symmetric, because all higher values of $l$ are
$(2l+1)$-fold degenerate, and its $\lambda$ must be real. We label its
mode function $f_{00{*}}$ and its eigenvalue $\lambda_{0{*}}$.

Close enough to the singularity, we can neglect all stable
perturbation modes. The unstable perturbation keeps growing however, and
eventually the solution deviates nonlinearly from the critical
solution.  The solution eventually forms a black hole for
$p>p_*$, but disperses for $p<p_*$. As $p\to p_*$ from either side,
the solution approximates the critical solution to ever smaller
scales, that is to ever larger $\tau$. This means that $C_{00{*}}(p)$ must
vanish precisely at $p=p_*$. Linearizing $C_{00{*}}$ around $p=p_*$, and
neglecting all decaying  perturbations, we have
\begin{equation}
Z\simeq Z_*(x) 
+ {\partial C_{00{*}}\over \partial p}(p_*)(p-p_*) e^{\lambda_{0{*}}\tau}
f_{00{*}}(x).
\end{equation}
For $p>p_*$, we define $\tau_p$ as a function of $p$ by
\begin{equation}
{\partial C_{00{*}}\over \partial p}(p_*)(p-p_*) e^{\lambda_{0{*}}\tau_p}
\equiv \epsilon,
\end{equation}
where $\epsilon$ is an arbitrary small constant. 

We now consider the Cauchy data $Z_0$ given by
\begin{equation}
Z(\tau_p,x,\theta,\varphi) \simeq Z_*(x) +
\epsilon f_{00{*}}(x) \equiv Z_0(x).
\end{equation}
We have chosen $\epsilon$ small enough that the growing mode has not
yet gone nonlinear (its exact value does not matter). For larger
$\tau$, the linearized ansatz (\ref{ansatz}) breaks down, and we know
only that the solution forms a black hole and settles down to
Schwarzschild. We have defined $\tau_p$ so that the rescaled variables
$Z$ are independent of $p$ in the Cauchy data $Z_{0}$. The field
equations themselves are scale-invariant (or asymptotically
scale-invariant on small scales).  As a result, there is only one
scale, $e^{-\tau_p}$, in the problem, and therefore the black hole
mass must be proportional to it. We have
\begin{equation}
\label{mass_scaling2}
M\simeq M_0 e^{-\tau_p} \simeq C (p-p_*)^\gamma, \qquad \gamma = {1\over
\lambda_{0{*}}},
\end{equation}
where $M_0$ is a universal constant, and $C$ is a
family-dependent constant absorbing $\partial C_{00{*}}/\partial
p(p_*)$.  This derivation of the critical scaling of the black hole
mass has been given before
\cite{KoikeHaraAdachi,Maison,HE2,Gundlach_Chop2}. Collapse simulations
show that the critical solution is an attractor, and the mass scaling
(\ref{mass_scaling2}) holds, even beyond the strict
applicability of the perturbation ansatz (\ref{ansatz}).

\paragraph{Angular momentum scaling}

Now we generalize the above discussion to initial data with
nonvanishing angular momentum. Our discussion holds strictly only for
initial data which are almost spherically symmetric, so that deviations from
spherical symmetry can be treated as linear perturbations throughout
the evolution. In particular this means that our results hold only for
black holes whose angular momentum is much smaller than their mass, so
that the final Kerr solution can be treated as a perturbation of
Schwarzschild. If the critical solution turned out to be an attractor
beyond the linear regime also for non-spherical initial data, our
discussion and results could then hold also for larger deviations from
spherical symmetry.

A Kerr black hole metric with small angular momentum ($L\ll M^2$) can
be written as a Schwarzschild metric with a linear perturbation
proportional to $L$. In fact, the gauge can be chosen so that the
shift component of the metric is
\begin{equation}
g_{t\varphi} = 2{L\over r}\sin\theta{\partial\over\partial\theta} P_1 
+ O(L^3/M^6),
\end{equation}
where $P_1$ is a Legendre polynomial, while all other metric
coefficients are those of the Schwarzschild metric in the usual
Schwarzschild coordinates, plus perturbations of $O(L^2/M^4)$
\cite{Gleiser_etal}. This result can be verified immediately by
expanding the Kerr solution in Boyer-Lindquist coordinates around
$L=0$.  The only perturbation to leading order in $L$ therefore has
$l=1$ angular dependence and odd parity in the Regge-Wheeler
notation. Here the orientation of the coordinate axes was chosen so
that the perturbation is purely $m=0$. For a general orientation of
$\vec L$ relative to the coordinate axes, $m=-1,0,1$ will all appear.
As a subdominant effect, we therefore include the odd $l=1$
perturbations in the data $Z_0$ even though they are subdominant, with
the aim of learning something about the behavior of the black hole
angular momentum near criticality.

We only keep the dominant odd $l=1$ mode, that is the one with the
largest (least negative) real part of $\lambda$, and denote it by
$\lambda_{1{*}}$. Generally, $\lambda_{1{*}}$ will not be real, but part
of a complex conjugate pair. $l=1$ is also threefold
degenerate, with mode functions $f_{1m{*}}$ for $m=-1,0,1$, or the three
directions in space. Keeping these modes results in a small perturbation
of the initial data $Z_0$:
\begin{equation}
Z(\tau_p,x,\theta,\varphi) \simeq Z_0(x) + e^{{\rm
Re}\lambda_{1{*}}\tau_p} \delta Z_0(\tau_p,x,\theta,\varphi),
\end{equation}
where
\begin{equation}
\delta Z_0 = {\rm Re} \sum_{m=-1}^1 
C_{1m{*}}(p_*) e^{i {\rm Im} \lambda_{1*} \tau_p} f_{1m{*}}(x,\theta,\varphi) 
\end{equation}
is periodic in $\tau_p$. As long as its amplitude $e^{{\rm
Re}\lambda_{1{*}}\tau_p}$ is small enough, the perturbation $\delta
Z_0$ evolves as a linear perturbation of the spacetime generated by
the spherically symmetric data $Z_0$ all way until the solution has
settled down to Schwarzschild with a small odd-parity $l=1$
perturbation that turns it into a slowly rotating Kerr solution.  By
the black hole no-hair theorem, no other perturbations (except an
electric charge of the black hole) can survive in the final
state. This justifies our considering odd $l=1$ perturbations but
neglecting $l>1$ and even perturbations altogether.  In the linear
approximation, the black hole angular momentum $L$ must, on average in
$\tau_p$, be proportional to $e^{{\rm
Re}\lambda_{1{*}}\tau_p}$. Because this is a dimensionless number, we
must have
\begin{equation}
{L\over M^2} \propto e^{{\rm Re}\lambda_{1{*}}\tau_p}
\end{equation}
for the overall scaling of $L$. But there is an interesting modulation
due to the fact that $\delta Z_0$ depends periodically on $\tau_p$, as
we shall see now.

By linearity, the presence of a perturbation $\epsilon\,{\rm Re}
f_{1m{*}}$, for some small $\epsilon$, in the initial data $Z_0$ must
give a rise to a small angular momentum component $L_z/M^2 = \epsilon A$
in the final black hole, while $\epsilon\,{\rm Im} f_{1m{*}}$ gives
rise to $L_z/M^2 = \epsilon B$. Similar proportionalities apply to ${\rm
Re} f_{1m{*}}$ and ${\rm Im} f_{1m{*}}$ for $m=\pm1$, and the $x$ and
$y$ components of the black hole angular momentum. We absorb the
universal, but unknown proportionality factors and the
family-dependent constants $C_{1m{*}}$ together into six new constants
$\vec A$ and $\vec B$. Putting all the factors together, we obtain the
following final result for the angular momentum vector $\vec L$ of the
black hole as a function of the initial data, for initial data near
the black hole threshold and near spherical symmetry:
\begin{equation}
\label{CSS_L}
\vec L = {\rm Re}\left[ (\vec A + i \vec B)
(p-p_*)^{\mu+i\omega} \right],
\end{equation}
where
\begin{equation}
\mu = {2 + {\rm Re}\lambda_{1{*}} \over - \lambda_{0{*}}}, \qquad
\omega = {{\rm Im} \lambda_{1{*}} \over - \lambda_{0{*}}},
\end{equation}
and $\vec A$ and $\vec B$ are six family-dependent constants. Note
that we always have $\mu> 2\gamma$, because ${\rm Re}\lambda_{1{*}}<0$
by the assumptions of this calculation.

This is our main result. As $p\to p_*$ from above in the initial data,
the angular momentum vector $\vec L$ of the final black hole rotates
in space as it decreases.  It describes a rapidly shrinking ellipse of
size $(p-p_*)^\mu$. This behavior formally resembles a damped
isotropic 3-dimensional harmonic oscillator, with $-\ln(p-p_*)$
playing the role of ``time'', $\omega$ the frequency and $\mu$ the
damping. Of course, $p-p_*$ is not in any sense a time, but a measure
of the distance, in phase space, of the initial data set labeled by
$p$ from the black hole threshold.  

\paragraph{The axisymmetric case}

The result (\ref{CSS_L}) is particularly surprising if we consider a
family of strictly axisymmetric initial data. Then $\vec L$ can only
point along the symmetry axis, and the ellipse degenerates into a
line:
\begin{equation}
\label{axisymmL}
L=L_z=(p-p_*)^\mu A\cos[\omega\ln(p-p_*)+c],
\end{equation}
where $A$ and $c$ are two family-dependent constants.  If
the family of axisymmetric initial data has angular momentum that does
not vanish at $p=p_*$, then the angular momentum of the black hole, as
a function of the initial data, alternates between parallel and
antiparallel to the angular momentum of the initial data.  The simple
power-law scaling $L_z=(p-p_*)^\mu$ one might naively expect would be
recovered only if $\lambda_{1{*}}$ was purely real. The oscillating
behavior is sketched in Fig. 1. It is perhaps as dramatic an
illustration of the ``forgetting'' of initial data in critical
collapse as the universality of the critical solution itself.

\paragraph{Discrete self-similarity}

Here we have considered the effects of a critical solution that is
CSS, both for simplicity of notation and because the one case that our
calculation is actually known to apply to (the perfect fluid) is
CSS. It is straightforward, however, to generalize our result for the
angular momentum scaling to hypothetical critical solutions that are
DSS instead of CSS.  The periodic dependence of the angular momentum
in CSS critical collapse is a simple cosine, because it derives from a
time-dependence $\exp\lambda_{1{*}}\tau$ of the leading $l=1$ odd
perturbation, with $\lambda_{1{*}}$ complex.  In DSS critical
collapse, the background critical solution itself is periodic in
$\tau$, in a nontrivial way, with a period $\Delta$. Therefore, one
would expect both kinds of periodicity to arise in $\vec L$. The
calculation bearing this out is a straightforward generalization of
the one for the CSS case, and here we only give its result. It is
\begin{equation}
\label{DSS_L}
\vec L = {\rm Re}\left\{{\cal M}[\ln(p-p_*)+c]\, 
(\vec A + i \vec B)\,
(p-p_*)^{\mu+i\omega} \right\},
\end{equation} 
where the only difference is the presence of the complex $3\times3$
matrix ${\cal M}$. It is universal and can in principle be calculated
from the mode functions $f_{1m*}$. In the CSS case it can be assumed
to be 1. It is periodic in its argument $\ln(p-p_*)$ with period
$\Delta/\gamma$. $c$ is a family-dependent constant. The role of
${\cal M}$ is similar to that of the fine structure of the black hole
mass scaling in the DSS case
\cite{Gundlach_Chop2,HodPiran_wiggle} (where the same parameter $c$
also appears).

\paragraph{Predictions for the $p=\rho/3$ perfect fluid}

We conclude this paper with a quantitative prediction, based on
numerical work reported elsewhere \cite{Gundlach_codim2}.  The
previously known critical solution (in spherical symmetry) for the
perfect fluid with equation of state $p=\rho/3$ \cite{EvansColeman}
was perturbed, with the result that, at least perturbatively around
spherical symmetry, it remains the critical solution in the full phase
space. The assumptions of the present paper therefore apply to this
matter model. The values $\lambda_{0{*}} \simeq 2.785$ and
$\lambda_{1{*}} \simeq -0.226
\pm 0.644i$ were found numerically. The value of
$\lambda_{0{*}}$ gives rise to a critical exponent for the black hole
mass $\gamma=1/\lambda_{0{*}}\simeq 0.359$, in agreement with the
value $0.3558$ given previously
\cite{KoikeHaraAdachi,Maison}. Therefore, we predict, for this matter
model, for initial data near spherical symmetry and near the black
hole threshold, the oscillatory behavior (\ref{CSS_L}), with $\mu
\simeq 0.799$ and $\omega \simeq 0.231$. For axisymmetric critical
collapse in particular (which may be simpler to investigate
numerically than the general case) this reduces to the scaling
(\ref{axisymmL}) with the same $\mu$ and $\omega$ (as $\lambda_{1{*}}$
is not purely real).




\begin{figure}
\label{Lzfig}
\epsfxsize=8cm
\epsffile{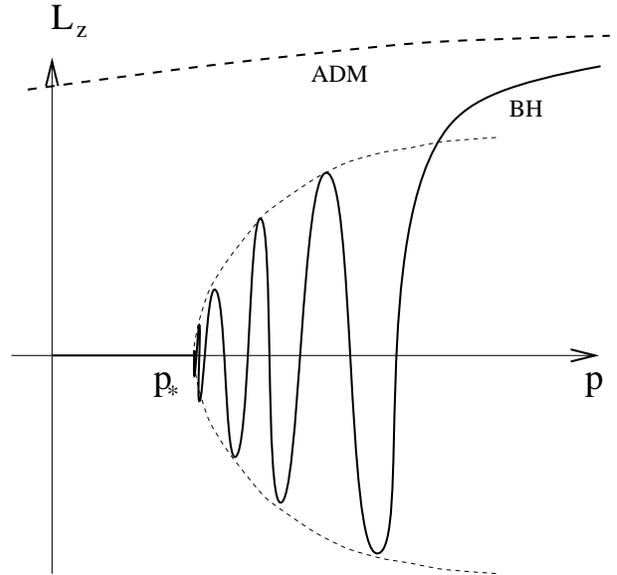}
\caption{Schematic plot of the ADM angular momentum (thick dashed
line) and the angular momentum of the black hole (thick full line) in
axisymmetric collapse in a one-parameter family of axisymmetric data
crossing the black hole threshold. $p$ is the parameter of the initial
data, and $p_*$ is its value at the black hole threshold. The envelope
(thin dashed line) is the overall power law. There is an infinite
number of oscillations as $p\to p_*$ from above.}
\end{figure}



\begin{references}

\bibitem{Choptuik} M. W. Choptuik, {Phys. Rev. Lett.} {\bf 70}, 9
(1993).

\bibitem{ChoptuikChmajBizon} M. W. Choptuik, T. Chmaj and P. Bizon, 
{Phys. Rev. Lett.} {\bf 77}, 424 (1996). 

\bibitem{BradyChambersGoncalves} P. R. Brady, C. M. Chambers,
S. M. C. V. Gon\c calves, Phys. Rev. D {\bf 56}, 6057 (1997).

\bibitem{Gundlach_critreview} C. Gundlach, Critical phenomena in
gravitational collapse, submitted to
Adv. Theor. Math. Phys., preprint gr-qc/9712084.

\bibitem{GundlachMartin} C. Gundlach and J. M. Mart\'\i n-Garc\'\i a,
{Phys. Rev.} D {\bf 54},  7353  (1996). 

\bibitem{HodPiran_charge} S. Hod and T. Piran, {Phys. Rev.} D {\bf
55}, 3485  (1997). 

\bibitem{Gundlach_codim} C. Gundlach, Nonspherical perturbations of
critical collapse and cosmic censorship, preprint gr-qc/9710066.  

\bibitem{KoikeHaraAdachi} T. Koike, T. Hara, and S. Adachi, 
{Phys. Rev. Lett.}  {\bf 74}, 5170 (1995).

\bibitem{Maison} D. Maison, {Phys. Lett.} B {\bf 366}, 82 (1996).

\bibitem{HE2} E. W. Hirschmann and D. M. Eardley, {Phys. Rev.} D
{\bf 52}, 5850 (1995).

\bibitem{Gundlach_Chop2} C. Gundlach,  {Phys. Rev.} D {\bf 55},
695 (1997).

\bibitem{Gleiser_etal} R. J. Gleiser, C. O. Nicasio, R. H. Price, and
J. Pullin, Evolving the Bowen-York initial data for spinning black
holes, preprint gr-qc/9710096.
 
\bibitem{HodPiran_wiggle} S. Hod and T. Piran, {Phys. Rev.} D {\bf
55}, 440 (1997).

\bibitem{Gundlach_codim2} C. Gundlach, in preparation, to be submitted
to Phys. Rev. D.  \bibitem{EvansColeman} C. R. Evans and
J. S. Coleman, {Phys. Rev. Lett.} {\bf 72}, 1782 (1994).

\end{references}
\end{document}